\documentclass[fleqn,10pt]{olplainarticle}

\usepackage{todonotes}
\usepackage{hyperref}
\usepackage[anythingbreaks]{breakurl}

\title{An Internet Voting System Fatally Flawed in Creative New Ways}

\author[1]{Andrew W.\ Appel}
\author[2]{Philip B.\ Stark}
\affil[1]{Princeton University}
\affil[2]{University of California}


\begin{abstract}
The recently published ``MERGE'' protocol is designed to be used in the prototype CAC-vote system.  
The voting kiosk and protocol 
transmit votes over the internet and then transmit voter-verifiable paper ballots through the mail.
In the MERGE protocol, the votes transmitted over the internet are used
to tabulate the results and determine the winners, but audits and recounts use the paper ballots that arrive in time.
The enunciated motivation for the protocol 
is to allow (electronic) votes from overseas military voters to be included in preliminary results before a (paper) ballot is received from the voter.  
MERGE contains interesting ideas that are not inherently unsound; but
to make the system trustworthy---to apply the MERGE protocol---would require major changes to the laws, practices, and technical and logistical abilities of U.S.\ election jurisdictions.
The gap between theory and practice is large and unbridgeable for the foreseeable future.  
Promoters of this research project at DARPA, the agency that sponsored the research, should acknowledge that MERGE is internet voting (election results rely on votes transmitted over the internet except in the event of a full hand count) and refrain from claiming that it could be a component of trustworthy elections without sweeping changes to election law and election administration throughout the U.S.
\end{abstract}

\begin{document}

\flushbottom
\maketitle

\thispagestyle{empty}

\section{Introduction}

Internet voting is ``the electronic return of voted ballots over the internet.'' \cite{acmusppc2010,vv23:iv,ncsl24:iv}
It is well understood that internet voting cannot be secured well enough to be used in public elections \cite{NAP25120}.  
Some ``end-to-end verified internet voting'' (E2E-VIV) protocols have been proposed; however, ``E2E-verifiability protocols are not, in and of themselves, sufficient to secure internet voting, even in theory.'' \cite[Section 5]{NAP25120}.  
Some of the problems with E2E-VIV are 
(1)~the voter and the local election official (LEO) must rely, for executing the complex cryptographic protocol, on computers that are themselves hackable; and (2)~voters must actively perform some parts of the protocol themselves, which is problematic given that most voters don't do the most basic review of machine-marked paper ballots \cite{demillo18,bernhard20}.

Recently a prototype internet voting system called ``CAC-Vote'' has been described \cite{adida24:merge,CACVoteFAQ,seck24,snyder24}.  
CAC-Vote comprises:
\begin{itemize}
    \item voting kiosks to be installed and used by U.S.\ military personnel and dependents at military bases remote from the home voting jurisdictions of those voters; 
    \item a computer system to be used by the local jurisdiction receiving the ballots from those kiosks;
    \item a \emph{cryptographic} protocol by which the voter's ballot is encoded and electronically transmitted, in addition to being printed on paper and mailed physically;
    \item a \emph{risk-limiting-audit} (RLA) protocol by which one could, in principle, detect whether hacking or other interference changed the electoral outcome and correct the outcome that occurred.
\end{itemize}

The cryptographic and RLA protocols are together called ``MERGE,'' an acronym for
``Matching Electronic Results with Genuine Evidence'' \cite{adida24:merge}.  
The ``electronic results'' are tallies that include votes transmitted over the internet; the ``genuine evidence'' is the human-readable paper ballots
that voters can hold, inspect, put in envelopes, and mail to election offices where local election workers can read and inspect them.

The design of CAC-Vote and its MERGE protocol make the standard assumption (in analyzing voting systems) that computers (such as the kiosk computer and the local election office's computer) are subject to wholesale hacking from remote locations; and the standard assumption that the U.S.\ mail and military mail systems for sending paper ballots are well protected and that any hacking of the paper trail would have to occur retail, one envelope at a time.  
So the problem that CAC-Vote and MERGE address is, ``can we count the electronic ballot right away, without waiting the (approximately) 5~days it takes an overseas priority-mail envelope to get to the LEO?''

One simple approach would be to match each paper ballot (if and when it arrives) with its cast-vote record (CVR), and ``uncount'' the electronic-ballot CVRs for paper ballots that don't arrive, or where the paper votes don't match the electronic votes, replacing them with the votes read from the paper ballot instead.  
But this is difficult to do without compromising the secret ballot.
So instead, the MERGE protocol 
uses the paper ballots 
in a way that does not connect the contents of the ballot to the particular voter (though MERGE acknowledges that ballot privacy may be compromised at some stage, and that it is unable to provide ``everlasting privacy"). 
The audit step of MERGE
is designed to
detect whether discrepancies between the electronic ballot CVRs and the paper ballots (including paper ballots that never arrived) 
altered the apparent outcome of the election---without necessarily matching each paper ballot to a specific electronic ballot.

There are other aspects to CAC-Vote system besides the MERGE protocol.  
A standard problem in any voting system is voter identity authentication.  To address this, CAC-Vote aspires to leverage U.S.\ military ID smart-cards, called "Common Access Cards'' or CAC \cite{CAC-manual}.  
And CAC-Vote solves another logistical problem:
U.S.\ law requires every one of those 9000 jurisdictions to 
provide a way for each overseas or remote military voter to obtain, through the internet, the \emph{unvoted} ballot in the right ballot style, to be printed, marked, and mailed---but not every soldier has easy access to a printer, and the CAC-Vote kiosk incorporates that printer.

Those are the things that CAC-Vote and MERGE are supposed to do.  
In this paper, we analyze the system and conclude:
\begin{itemize}
\item Sending an untrustworthy electronic vote to be counted, backed up by a paper ballot that's the genuine evidence---but that will not be counted unless there is a binding recount with suitable rules---is a solution in search of a problem; it is unnecessary.
\item The MERGE protocol is so mismatched to U.S.\ states' election laws and practices that it cannot be implemented.
\item MERGE does not contain adequate resolution procedures for (entirely foreseeable) things that can go wrong.
\item DARPA has published a FAQ about the project which contains unsupported statements and inaccurate claims, and assertions that contradict the paper published by the project's scientists. 
\item The CAC-Vote kiosk concept has interesting and useful innovations concerning voter authentication, convenience, ballot return, that could be useful \emph{if the electronic ballot return feature is removed and the MERGE protocol is not used}.
\item The CAC-Vote kiosk could have the capabilities to usefully address other problems with
paper ballot tracking.
\item Given the impracticality of deploying MERGE, the Defense Advanced Research Projects Agency (DARPA) and its staff should stop promoting this as a solution for military voters.
\item Given that the system uses the electronic votes transmitted over the internet to determine who won and that the mailed paper ballots play no role in determining who won unless there is a full manual recount of the mailed votes,  DARPA should stop claiming that this is ``not internet voting.''

\end{itemize}

\section{How CAC-Vote/MERGE works}

Our description of the system is based on the published paper \cite{adida24:merge}, the FAQ \cite{CACVoteFAQ}, and interviews with Ben Adida (president of Voting.Works, principal investigator of the DARPA-funded research project),
Dan Wallach (program manager at DARPA, which funded this research), and
Vanessa Teague (cryptographer who led the design of the MERGE protocol).

The military voter inserts their CAC card into the CAC-Vote kiosk computer.  
This authenticates the voter to the system, based on cryptographic signatures in the CAC card and 
the Department of Defense's mechanisms for identity verification in issuing CAC cards.
Somehow, the kiosk determines the ``Local Counting Center,'' i.e., internet address
of the state or county
election office of the voter's home jurisdiction---though the MERGE paper
does not explain how this is done.
Then, ``Observers and officials in each Local Counting Center know
the CAC IDs of the voters they should include, and can validate
CAC certificates''  \cite[\S1.1]{adida24:merge}.  This is stated as a fact (though 
perhaps it was intended to be read as an assumption) but
it is certainly not the case in 2024 at most local election offices.
It is not at all clear how practical or costly it would be to build and maintain a database of the 9000 different U.S.\ voting jurisdictions and integrate the CAC-vote receiving software into each of them; 
nor what entity would have the authority, responsibility, and resources to do so; nor if all 9000 jurisdictions would cooperate in its deployment;
and it is not clear whether the mechanisms to do this are funded as part of this DARPA-sponsored research project or are left as future work.

Assuming that the local-election-office computer could look up the voter's CAC number and connect
it to a voter-registration record, that would determine the \emph{ballot style}
the voter should receive---the set of election contests in which a voter registered at this 
address is eligible to vote in.
The kiosk obtains the \emph{ballot definition} corresponding to this ballot style, and presents the ballot to the voter on a touchscreen.  
The voter indicates their selections, and when finished, the kiosk prints out a human-readable paper ballot and a mailing envelope on which the jurisdiction's election-office address is printed.  
Perhaps 7\% of the time, the voter reviews the ballot selections carefully to make sure that they are the same as what they selected on the screen; the rest of the time, the voter does not bother 
or does not look long enough even to count the contests, much less verify the selections
\cite{bernhard20,demillo18,haynes21}. 
Then the voter puts the printed ballot into the envelope (or double envelope if their state requires it), signs the envelope, seals the envelope, and gives it to their military base's Voting Assistance Officer.  
Like any remote military absentee paper ballot, this envelope is eligible for an \emph{11-DOD} Priority Mail label, at no cost to the voter, which (like other U.S.\ Postal Service Priority Mail) has a bar code that allows the voter to track the envelope to its destination.  
The Department of Defense, through its \emph{Federal Voting Assistance Program}, pays particular attention to its system for expedited shipment of military Priority Mail, especially overseas absentee ballots, which are received in Chicago and distributed from there to the continental United States.  
End-to-end transit time for such envelopes has been measured to be about five days \cite{MPSA2010}.

Meanwhile, in addition to printing out the paper ballot, the CAC-Vote kiosk encrypts the voter's selections 
and transmits them over the internet to the local election office.  
The encryption is done in a way that preserves the secret ballot, that is, disassociates the voter's identity from the voter's choices.  
The local jurisdiction's computers counts these votes immediately, without waiting for the paper ballot to arrive.

An important assumption behind CAC-Vote/MERGE is that computers can be hacked, including the kiosk and the LEO's computer that receives the electronic ballot; but the paper ballot (that the voter can hold in his hands and verify) is reliably sent through the military mail system and U.S.\ Postal Service, in the sense that it may or may not arrive in time, or at all, but if it arrives then it is genuine and has not been substituted or modified.

The consensus of election security experts is that electronically
returned ballots are vulnerable to large-scale remote attacks
and manipulation, but physically mailed ballots are relatively
secure. 
Furthermore, the voter can see the votes marked in
the paper ballot, but has no way of knowing what
was transmitted digitally.
Therefore, 
if an electronic ballot arrives and the corresponding paper ballot either disagrees with the electronic ballot or does not arrive at all, then the electronic ballot should not be included in the tally---because it may be the result of fraudulent computer hacking.  

An important assumption behind CAC-Vote/MERGE is that, although the paper ballot may not arrive by election day (when votes are counted and unofficial results are announced), the paper ballot will arrive before the deadline for auditing and certifying official election results.  
The \emph{intent} of the MERGE protocol is to
count the electronic ballot early (on election day), but permit the ballot
to contribute to the official election outcome 
only if the paper ballot arrives before
before the statutory deadline for receiving absentee ballots
\emph{and} the paper ballot agrees with the electronic ballot.

You might imagine that if the paper ballot fails to arrive or if it disagrees with the electronic ballot, the electronic ballot is removed from the count.  
But that would not be easy to do, especially while also preserving the privacy of the voter's secret ballot. 
So MERGE is a clever combination of cryptographic protocol
and statistical risk-limiting audit (RLA) protocol that tries to ensure that
\emph{even though you might count an electronic ballot whose paper mismatches or doesn't arrive, you are unlikely to do so if 
counting all such electronic ballots would change who wins the election}.  

RLAs are normally used with hand-marked, optically scanned paper ballots.  
RLAs manually examine a random sample of the paper ballots. 
With high assurance, if a hack or error caused the election
system to claim the wrong outcome---compared to what
the voters actually chose---then the RLA will lead to a recount of the paper ballots to correct the outcome before the outcome is certified.

The MERGE protocol requires the local jurisdiction to perform a properly designed and executed RLA that includes all ballots cast in every election contest---not only the CAC-Vote ballots.
MERGE provides mechanisms to include the CAC-Vote ballots in the RLA.
For 
example, if an electronic vote is counted (present in the electronic CVR file) but the paper ballot never arrives, the MERGE/RLA accounts for that \emph{by assuming, for audit purposes, that the ballot had a valid vote for every losing candidate in every election on the ballot}.  
This ensures that if computer hacking of the CAC-Vote system could have altered the outcome of the election, then the RLA has a large chance of
calling for a full recount.

That's the theory.  
But in practice, there are so many obstacles to the implementation and proper operation of this system that it is 
impossible to deploy, and will be for years to come, in any U.S.\ jurisdiction.

\section{Why CAC-Vote is insecure in practice}

Then entire MERGE protocol, designed to reconcile the paper ballots with the electronic ballots, depends on its integration into a binding 
risk-limiting audit (RLA) that the local jurisdiction performs.  
Without the RLA, there is no mechanism to check that the voter's intent is the vote that counts.

But only 5\% of U.S.\ voters live in states that have risk-limiting audits that are binding on election outcomes:  Colorado, Rhode Island, and Virginia.\footnote{See \url{https://verifiedvoting.org/auditlaws/}}
And in those states, only some contests are subject to an RLA.\footnote{%
In Colorado, the Secretary of State selects the contests to audit and may select those contests to minimize the workload or for any other reason.
\url{https://www.sos.state.co.us/pubs/info\_center/laws/Title1/Title1Article7.html\#a1-7-515},
\url{https://www.sos.state.co.us/pubs/rule\_making/CurrentRules/8CCR1505-1/Rule25.pdf}
In recent years, Colorado has audited just two contests in each county.
In Rhode Island, the state board of elections determines which contests to audit. 
\url{https://casetext.com/statute/general-laws-of-rhode-island/title-17-elections/chapter-17-19-conduct-of-election-and-voting-equipment-and-supplies/section-17-19-374-post-election-audits}
In Virginia, RLAs are required for only one or two contests, and contests with small margins are specifically excluded.
\url{https://law.lis.virginia.gov/vacode/24.2-671.2/}.
}
Even in those states, the security of CAC-Vote 
would depend on changes in state law to integrate its complex
protocol and to require an RLA of \emph{every contest in every election}, regardless of the reported margin and anticipated workload.
In any other state,  CAC-Vote can be no more secure than any other form of internet voting.  

The lack of RLAs is not the only problem, however.  
There are quite a few ways that hacking the CAC-Vote computers could allow criminals to get away with changing election outcomes.  
\begin{enumerate}
    \item As noted, inadequate RLAs in every state mean that CAC-Vote is just a plain-old internet voting system with no ``genuine evidence;'' until the states have real RLAs of every contest in every election, MERGE is just an interesting research result, not a practical system.  
    In general, the MERGE paper is  careful in its claims and  restricts itself to explaining this research result; but public statements by DARPA officials and Voting.Works executives make the false claim that MERGE is not internet voting and promise more than can be delivered.
    \item A hacked kiosk could put fraudulent votes (not those that the voter indicated on the touchscreen) into both the electronic ballot and the paper ballot. 
    Studies show that most voters who use touchscreeen ballot-marking devices (BMDs) barely glance at the paper ballot that gets printed out, even though the purpose of that ballot is for verification \cite{haynes21,demillo18}, and if the BMD
    deliberately changes the vote in one contest, only 7\% of voters would notice
    \cite{bernhard20}.  
    For those few voters that do check their printed ballots, 
    assuming the CAC-Vote kiosk has some provision for voiding a ballot and starting over, that just means that a hacked CAC-Vote could steal only 93\% of the votes that it tries to steal.  \emph{The MERGE protocol cannot recover the original voter-indicated vote in this case.}
    \item A hacked kiosk could put the wrong mailing address onto the envelope it prints out---if the votes on that ballot are not for the candidate that the hacker prefers.\footnote{Our analysis is of Version~2 of the MERGE paper on the arXiv.  We are told that Version~3 may acknowledge this point.}  
    Then the paper ballot will fail to arrive, which is supposed to mean (according to the MERGE protocol) that the voter's ballot can't affect the election outcome.
    \item A hacked kiosk could give incorrect instructions for how to handle and verify the ballot \cite{appel23:swiss}.  
    Therefore, any part of the protocol that relies on actions by the voter may fail to be executed correctly.  
    \item CAC-Vote prints a sticker containing a digital signature, which the voter affixes to the envelope and which can authenticate the paper ballot to the election office.  
    This is supposed to improve security.  Invalid 
    signature stickers are cause for rejection
    at the election office that receives the
    paper ballot.  
    But a hacked CAC-Vote could deliberately produce invalid signatures on ballots with votes for candidate Smith, but put valid signatures on votes for Jones. That would turn this feature into a vulnerability, not a strength.    
    The MERGE protocol tries to address this problem by ``Each printed signature on the sticker is verified before
sending, either by the voter or by some other trustworthy
assistant at the Remote Voting Center.''
But the voter or the Voting Assistance Officer
need to use a computer for that purpose; what
protects \emph{that} computer from hacking?  
    \item The MERGE protocol's cryptographic checking of received electronic ballots, and the protocol's handling of the audits of received paper ballots, must be implemented on a computer system because local election officials cannot execute cryptographic protocols in their heads.  
    Very likely the software will be provided by the same company that provides CAC-Vote kiosks.  
    A criminal who hacks one could also hack the other, corrupting the MERGE protocol and rendering it ineffective.  
    To be fair, this cryptographic checking can be performed independently on any observer's computer, using any observer's installation of the open-source software; but as a practical matter it implausible that much independent checking will occur.
\end{enumerate}
Aside from hacking, it may be quite difficult to integrate CAC-Vote/MERGE into state
procedures and laws.
No state that allows ballots to be returned electronically has any state law, rule or regulation that would allow or direct a paper ballot received after Election Day to be used in an audit or recount. In addition, many states allow military ballots received after Election Day to be counted, making CAC-Vote's electronic ballot unnecessary. 
\begin{enumerate}
    \item[7.] More than half the states \cite[section IX]{appel23:unh} already permit military voters to return ballots by internet---which, of course, is quite insecure---and the way many states handle incoming electronic ballots is to print them out onto paper optical-scan ballots, or mark the ballots by hand onto optical-scan ballots, for counting in the regular paper-ballot stream of absentee ballots.  
    If the jurisdiction handles CAC-Vote electronic ballots this way, then there would be \emph{two} paper ballots for each CAC-Vote voter.  
    As a practical matter, there may not be a way to remove the remade-from-electronic ballot from the ballot batches, in a recount.  That would make it impossible act meaningfully on indications from the RLA component of MERGE protocol.
    \item[8.] In most states, rules for counting votes are enacted into state law by the legislature.  
    Under the laws of most states, if the MERGE 
    RLA detected that a previously counted ballot should not be counted, it would be illegal to properly act on this information \cite{VV-ALD}.  
    Therefore, in most states, CAC-Vote/MERGE could not be used in the way intended without  changing state laws.
\end{enumerate}
Therefore, the temptation would be to use the electronic ballot return part of CAC-Vote without the MERGE-RLA safeguards.  
But then CAC-Vote is as insecure as other internet voting systems.  
And this is the expected case, for the reasons we have explained.

\section{Voters who don't follow instructions}

Section~5 of the MERGE paper is a security analysis of the combined cryptographic/RLA protocol, intended to demonstrate that the security of
MERGE is no worse than the security of a plain RLA using ordinary 
hand-marked paper ballots.
In this security analysis, it is assumed that a certain percentage
of voters do not ``follow instructions'' for verifying their votes.
For MERGE, the ``instructions'' include checking cryptographic
signatures, looking up those signatures on a public bulletin board
a few days after casting the vote, and checking that the printed paper
ballot has the same votes that the voter indicated on the touchscreen.

In a conventional RLA, one could also assume that some percentage of
voters do not ``follow instructions'' for verifying their vote.
In the case of a BMD-printed paper ballot, those ``instructions'' are
to review, contest by contest through the paper ballot, that 
the printed votes are the same as those that the voter had
selected on the touchscreen.   Indeed, in such an RLA, to the
extent that some fraction of voters do not ``follow instructions''
for verification, the effectiveness of the audit is compromised.

The percentage of voters who ``follow instructions''
using conventional BMDs has been measured at 7\%. 
That is, 93\% of voters who were 
(deliberately) given a printed ballot with one contest mismarked,
did not bring this to anyone's attention \cite{bernhard20}.  
This is the very
serious flaw with BMD-printed ballots: RLAs cannot be effective
because there is no trustworthy record
of the voters' expressed preferences.

Checking a paper printout that's already in your hands is at least fairly
intuitive, and only 7\% of voters do it well.  
The fraction of voters who follow the more difficult and unintuitive
digital-signature-verification and bulletin-board lookup instructions
must surely be smaller.
In the MERGE paper's security analysis, the fraction of
voters who do not ``follow instructions'' is left unquantified;
this is a serious omission, since surely if the number is extreme, it must undermine the security of the protocol.

The MERGE paper says that the MERGE protocol can be used with 
hand-marked paper ballots, scanned at the kiosk.  
This mitigates the
possibility that the voter will fail to check that the paper-marked ballots
match the votes they had indicated on the touch screen, because there are no
votes indicated on the touch screen.  But the other checks 
(digital signature and bulletin board) are needed regardless.

Now, a careful analysis of the protocol might find that failure to carefully review the BMD-printed paper ballot would have different consequences, for security and 
accuracy, than failure to check the digital signature or the bulletin board.
But CAC-Vote should not be considered for deployment until such an analysis is done.
\section{Ethics Considerations}
The MERGE paper makes the following disclaimer:
\begin{quotation}
\noindent\textbf{Ethics considerations.}
Work on voting always requires significant ethics considerations.
This protocol is designed for practical use, so we have
to consider carefully how it might be used in practice. 
The
main ethical consideration---which has been raised by informal
reviews we sought before formal submission---is that we
cannot guarantee the protocol is run in the conservative way
we describe here. 
People do not always follow instructions,
and might be tempted (or legally required) to count electronic votes without running
the RLA in the rigorous way we have specified. 
We
acknowledge this risk, but feel that this applies to any security
protocol. 
We have been careful to specify as clearly as possible
the procedural supports necessary to run this protocol securely.
\cite{adida24:merge}
\end{quotation}

This is a reasonable disclaimer to accompany a scientific paper that describes a novel cryptographic protocol.  
But in the
context of promoting new voting products to the public and to election officials, those disclaimers have not been made where they should have been.

The DARPA program manager who oversees the funding of the
CAC-Vote project and the principal investigator who leads
the project have been showing the system to state 
election directors and all but promising to election
officials and to military voters that the system will be 
ready to deploy in 2025.

In the newspaper \emph{Stars and Stripes}, July 2024:
\begin{quotation}
VotingWorks unveiled an early-stage version of its deployable voting machine in February at the National Association of State Election Directors conference in Washington, D.C. The machine aims to allow service members around the world, even at remote locations, to transmit a signed, encrypted digital ballot to their home precinct for tallying on Election Day. \ldots

[Ben] Adida [President of VotingWorks] said the machines will not be ready for use in this year's upcoming presidential election in November, but the company is hopeful to deploy them next year. 
``We are exploring possible opportunities for a pilot election in early-mid 2025.''\quad \cite{snyder24}
\end{quotation}

In the newspaper \emph{Military Times}, February 2024:
\begin{quotation}
Dan Wallach, program manager for the DARPA Innovation Office, said [that the year] 2025, after the close of the 2024 presidential election cycle, will present more opportunities for trial and experimentation.

``You don’t roll out new things in a high turnout, high-stakes election,” he said. “When we’re voting for the dog-catcher, that’s when you try the new stuff.''
\quad  \cite{seck24}
\end{quotation}
In support of CAC-vote, DARPA published an FAQ \cite{CACVoteFAQ} which includes several inaccurate statements and contradictory claims: 
\begin{quotation}
\noindent Q. Is this Internet Voting?\newline
A. No. \ldots\newline
\emph{[For the reasons stated above, CAC-vote is internet voting: that claim is false.]}
~\newline
Q: Will this be used for the 2024 elections?\newline
A: No.\newline
~\newline
Q: Will overseas military voters get a paper ballot?\newline
A: Yes. Overseas military voters using the system will have a paper ballot that they will then
mail back to election officials. At the same time, an encrypted ballot will be returned digitally
to election officials.\newline
~\newline
Q: What will be the official ballot, the paper or the digital ballot?\newline
A: That will be a policy decision that will be left up to each election jurisdiction based on their
laws and regulations.\newline
\emph{[Here DARPA acknowledges that the paper ballot may not be counted as the official ballot for recounts and audits, directly contradicting its rationale for claiming CAC-vote is not internet voting.]}
~\newline
Q: Who will count the ballots?\newline
A: There will be no change to the way ballots are counted today. Ballots will be counted by
the local election official in the state where the overseas military voter is registered to vote.\newline
\emph{[But the MERGE protocol counts votes from the mixnet of ballots; quite a change!]}\newline
Q: Has/will this system be tested/certified?\newline
A: The system that is the result of this research project will not be certified by the Election
Assistance Commission since it does not tally ballots and is only a ballot marking and
transmission system.\newline
\emph{[Although the kiosk does not tally votes, the mixnet component of CAC-Vote does tally votes.]}
\end{quotation}

\emph{CAC-Vote \textbf{is} internet voting.}
U.S.\ State Secretaries of State are the ones who make (or influence) important decisions about whether to permit internet voting in their states' elections.  
Promoting a system to them with the implication that the Defense Department endorses it as secure does not comport with the ``ethics considerations'' alluded to in the scientific paper.

Claiming that ``there will be no change to the way ballots are counted'' when the protocol expressly involves counting encrypted selections transmitted over the internet, and sidestepping the very difficult legal question of what will be the official ballot (the paper or the digital one), is a serious misrepresentation.  
In fact, it is the inherent difficulties in that very question---what \emph{is} the official ballot, in the MERGE protocol---that is at the heart of the problem.  For, unfortunately, the electronic ballot is counted but not trustworthy, and the paper ballot is not actually counted---it is reserved for audits and recount, which will likely never occur.
Even if there is an RLA as the protocol requires, the final results are based on the electronically transmitted votes \emph{unless the RLA leads to a full recount of the votes on the paper ballots}.

\subsection*{Why does it matter?}

The Department of Defense has a good record of prudence and public review when it comes to internet voting.  
In the 2002 National Defense Authorization Act (NDAA) Congress directed the DoD's
Federal Voting Assistance Program (FVAP) to develop an online ballot return system for military and overseas voters.   
DOD developed a Secure Electronic Voting and Registration Experiment (SERVE), intended to be deployed in the 2004 general elections. 
Before it was deployed, at the request of the DOD, independent security researchers examined the security of the SERVE system and found that it could not reliably secure ballots cast over the internet. Researchers concluded that the insurmountable challenge to securing an online voting system lies in the fundamentally insecure nature of the internet: 
\begin{quotation}
We do not believe that a differently constituted project could do any better job than the current team. 
The real barrier to success is not a lack of vision, skill, resources, or dedication; it is the fact that, given current Internet 
and PC security technology, the FVAP has taken on an impossible task. \cite{jefferson04}
\end{quotation}
Based on this science, the U.S. Congress and the Department of Defense have been careful and explicit about not
promoting internet voting systems.  
Congress followed the science, abandoned SERVE, and repealed a funding authorization to pursue internet voting.\footnote{This repeal was part of The Ronald W. Reagan Defense Authorization Act for FY2005 (P.L. 108-375).}  DoD
concentrated on improving the accessibility and delivery of remote absentee paper ballots.
DoD has explicitly prohibited the use of certain
federal funds for developing internet voting systems.  

It may be entirely appropriate
for DARPA to fund basic-science research projects such as the development of a 
combined cryptographic-and-auditing protocol that might someday mitigate some
of the risks of internet voting.  But it is inappropriate to make promises
to the Secretaries of State of the 50 states that a DARPA-endorsed internet voting system 
will soon be available. 

\section{What a CAC-based voting kiosk could be useful for}

A February 2024 article in \emph{Military Times} described the motivations behind CAC-Vote:
\begin{quotation}
Some 54\% of surveyed military members who wanted to vote but didn't\footnote{For context: in 2020, 49\% of military members voted, 21\%  wanted to vote but didn't complete the process, and 30\% didn't want to vote \cite{fvap-adm-2020}.  So 54\% of 21\% = 11\% had sufficient trouble requesting an absentee ballot that it hindered them from voting.} say they had trouble requesting an absentee ballot, and another 43\% said the ballot never arrived at all. 
Other reasons for not voting included a too-complicated voting process, difficulty with the mailing system, and trouble accessing their state’s election website. 

The technology proposal from VotingWorks would change the game with portable one-stop voting stations, compact enough to fit in a big suitcase, that print a paper ballot just like the one corresponding to a service member’s local election and also generate a label for mailing that ballot to the correct polling center, all the way up until the night of the election.  \cite{seck24}
\end{quotation}
Every one of those issues could be addressed by a kiosk that omits the electronic-ballot-return feature (and the MERGE protocol).  
We suggest that such a kiosk could accomplish more than that.

\begin{enumerate}
\item First, ditch the electronic ballot return, which is not needed for any of the soldiers' concerns listed above.
\item Ditch the touch-screen ballot-marking aspect and print a hand-markable paper ballot, to avoid the well known problems with the security of BMDs \cite{appel20:bmd,starkXie22}.
\item Focus on how to deploy CAC-based lookup of the appropriate local election website, which is not addressed in the paper; and within that website, how to connect CAC-ID to the voter registration record.  Both of these would be useful to voters.
\item Keep the print-envelope-address feature.  
In combination with the adjustment of printing a hand-markable paper ballot, this mitigates a security vulnerability: if a hacked kiosk wants to print a wrong address on some labels, it now must do so \emph{without} first seeing what votes are marked on the ballot.
\item Add a tracking number, or better yet, scan the tracking barcode on the 11-DOD label, for reasons we discuss below.
\end{enumerate}

The Department of Defense and the U.S.\ Congress have paid attention to the timely return of ballots from soldiers stationed remote from their voting jurisdictions.  The MOVE Act of 2010 requires jurisdictions to allow such voters to request absentee ballots by internet and even to request that the PDF of their (unvoted) ballot be available by internet, so that the voter can print it, mark it with a pen, and mail it back.  
Cybersecurity experts generally consider this use of the internet safely implementable.
DoD and Congress have also paid attention to the timely return of paper ballots, via the 11-DOD Priority Mail label discussed above.  
And the DoD Inspector General makes regular reports to Congress regarding the effectiveness of these and other programs associated with the Federal Voting Assistance Program (FVAP) of the DoD.

But there have not been regular assessments of how long it takes a paper ballot to get from the hands of the overseas soldier to the local election office.  
The most recent report we could find is from 2011 \cite{MPSA2010}.  
We conjecture that studies are rare because
mail latency is hard to measure.  
A system of standardized voting kiosks, such as CAC-Vote, could help.  
CAC-Vote could, for each paper ballot that it prints, record when it was printed,
track it through the postal system (using the Priority Mail barcode) all the way to the local election office, and aggregate statistics for each military base and all military bases.

Each military base has its own logistical issues with mail service, depending on its remoteness; each base's Voting Assistance Officer may have other duties or travels that interfere with timely collection of ballots and getting them into the mail system.  
Delays resulting from these issues may be excusable, but at least they should be measured.  
A CAC-Vote system could measure the delay between printing the ballot and its first appearance in the Priority Mail tracking system.

\section{Conclusion}

The recent history of internet voting ``solutions'' is that, 
one after another, they are deployed and then found insecure:
see ``Is Internet Voting Trustworthy?'' (section VI) for details \cite{appel23:unh}
of the D.C. system (insecure), the Estonian system (insecure), the Australian (insecure), 
the Swiss (insecure), Voatz (insecure), 
and Democracy Live (insecure).
Let's continue the research.
MERGE is an interesting contribution to the
research literature.  
But let's not pretend that any of these systems is even close to ready for deployment
before it is evaluated thoroughly in the scientific literature.

\subsection*{Acknowledgments}
Susan Greenhalgh, David Jefferson, Josh Benaloh, and Ron Rivest contributed to discussions that led to this article.  
We thank Ben Adida, Vanessa Teague, and Dan Wallach for their willingness to 
discuss the CAC-Vote system with us.
\bibliographystyle{plainurl}
\bibliography{voting}

\end{document}